\documentclass[aps,twocolumn,showpacs]{revtex4}
\usepackage{amsmath}
\usepackage{epsfig}

\begin{document}

\title{Can $\Omega(2012)$ be explained as a molecular state?}

\author{Xuejie Liu$^1$}\email[E-mail: ]{1830592517@qq.com}
\author{Hongxia Huang$^2$}
\email[E-mail: ]{hxhuang@njnu.edu.cn (Corresponding author)}
\author{Jialun Ping$^2$}
\email[E-mail: ]{jlping@njnu.edu.cn (Corresponding author)}
\author{Dianyong Chen$^1$}\email[E-mail: ]{chendy@seu.edu.cn}
\affiliation{$^1$School of Physics, Southeast University, Nanjing 210094, P. R. China}
\affiliation{$^2$Department of Physics, Nanjing Normal University, Nanjing 210023, P.R. China}

\begin{abstract}
We conduct a dynamical calculation of pentaquark systems with quark contents $sssu\bar{u}$ in the framework of two quark models: the chiral quark model(ChQM) and quark delocalization color screening model(QDCSM). The effective potentials between baryon and meson clusters are given, and the possible bound states are also investigated. Besides, the study of the scattering process of the open channels is also performed to look for any resonance state. The results show that the $\Omega(2012)$ is not suitable for the interpretation as a $\Xi^{*} \bar{K}$ molecular state in present quark models. Two resonance states: the $\Xi^{*}\bar{K}^{*}$ with $IJ^{P}=0\frac{3}{2}^{-}$ ($M=2328\sim2374$ MeV, $\Gamma=57\sim65.5$ MeV) and $IJ^{P}=1\frac{3}{2}^{-}$ ($M=2341\sim2386$ MeV, $\Gamma=31.5\sim100$ MeV) are obtained in both QDCSM and ChQM, which indicates that both of these two states are more possible to be existed and worthy of being searched by future experiments.
\end{abstract}

\maketitle

\setcounter{totalnumber}{5}

\section{\label{sec:introduction}Introduction}
The possible new resonance state is always an interesting topic in both experimental and theoretical physics, and it can help us to catch more details about the hadron-hadron and quark-quark interaction. The QCD theory has only predicted the existence of colorless hadronic states including two and three quarks. The hadrons which have more than three quarks lie outside the original quark model. These particals are the so-called exotic hadrons. The quantum numbers of these hadrons are unexplainable with the typical two and three quark bound states. Although the interaction fails in the description of many resonances, an example of the resonance better described by other approaches is the $\Lambda(1405)$ which can be interpreted as a meson baryon system~\cite{1405.1,1405.2,1405.3,1405.4} or the Roper resonance~\cite{Roper}.

In 2018, the Belle Collaboration searched for excited $\Omega^{-}$ decaying into $\Xi_{0} K^{-}$ and $\Xi^{-} K^{0}_{s}$ using $\Upsilon(1S)$, $\Upsilon(2S)$ and $\Upsilon(3S)$ date samples with integrated luminosities of $5.7$, $24.9$, and $2.9$ $fb^{-1}$, respectively~\cite{omeiga2012.2018}. A signal peak around $2.012$ GeV is seen with a signal significance of 8.3$\sigma$. This new resonance is denoted as $\Omega(2012)$, with a mass of $2012.4\pm0.7(stat.)\pm0.6(syst.)$ MeV$/c^{2}$ and a width of $6.4^{+2.5}_{-2.0}(stat.)\pm1.6(syst.)$ MeV. As we know that the ground $\Omega$ hyperon is a member of baryon decuplet in the quark model, which was unambiguously discovered in both production and decay at BNL about half century ago~\cite{omeiga1672}. Three resonances: $\Omega(2250)$, $\Omega(2380)$ and $\Omega(2470)$ are also listed in the Particle Data Group $(PDG)$~\cite{PDG} with three- or two-star ratings, and their nature are still uncertain. So in experiments there are only a few information of the excited $\Omega$ baryons. The discovery of the $\Omega(2012)$ state offers a great opportunity to study the hadron spectrum and understand the strong interactions.

Actually, before the Belle collaboration's report, there were some theoretical studies of the $\Omega$ excited states. Refs.~\cite{befor.2012.1,befor.2012.2} investigated the $\Omega$ excited states by using the chiral unitary approach where the coupled channels interactions of the $\Xi^{*}\bar{K}$ and $\Omega\eta$ were taken into account. In Ref~\cite{befor.2012.3}, a dynamically generated $\Omega$ state with $J^{P}=\frac{3}{2}^{-}$ was obtained with the mass around $1800$ MeV, which is in agreement with the predictions of the five-quark model. In addition, the $\Omega$ excited states were also investigated in classical quark models~\cite{befor_model1,befor_model2,befor_model3,befor_model4}.

The Belle collaboration's observation of the $\Omega(2012)$ also prompted a lot of theoretical work on the issue, with pictures inspired by quark models and also molecular pictures based on the meson-baryon interaction. In various quark models, the masses of the first orbital excitations of $\Omega$ states was deemed to $\Omega(2012)$. In the chiral model, through the study of the strong decays of $\Omega(2012)$, it could be assigned to the spin-parity $J^{P}=\frac{3}{2}^{-}$ $1P$ wave state~\cite{model1}. In Refs.~\cite{model2,model3}, the mass and the two-body strong decays of the $\Omega(2012)$ state were studied by the QCD sum rule method and the results shown that the $\Omega(2012)$ could be interpreted as $1P$ orbital excitation of the ground state $\Omega$ baryon with the quantum number $J^{P}=\frac{3}{2}^{-}$. However, some work investigated $\Omega(2012)$ as the molecular state. Since its mass is very close to $\Xi^{*} \bar{K}$ threshold, some work considered it as a $J^{P}=\frac{3}{2}^{-}$ $\Xi^{*}\bar{K}$ molecular state~\cite{model4,model5,model6}. In Ref.~\cite{model7}, the study of the $\Xi \bar{K}$ decay mode of the newly observed $\Omega(2012)$ indicted that the $\Omega(2012)$ was a dynamically generated state with spin parity $J^{P}=\frac{3}{2}^{-}$ from the $S-$wave coupled channel interactions of $\Xi \bar{K}$ and $\Omega\eta$.

However, it was found that there is no significant signals for the $\Omega(2012)\rightarrow \Xi^{*}\bar{K}\rightarrow\bar{K}\pi\Xi$ in a very recent measurement of the Belle collaboration~\cite{belle(2019)}. It is in sharp tension with the prediction of the S-wave $\Xi\bar{K}$ molecule assignment for $\Omega(2012)$, so
this result strongly disfavors the molecular interpretation. Later on, based on the experiment results, a lot of work reinterprets the $\Omega(2012)$ state.
In Ref.~\cite{model8}, the $\Omega(2012)$ was explained as the $P-$wave molecule assignments, and can be interpreted as the $\frac{1}{2}^{+}$ or $\frac{3}{2}^{+}$ $\Xi \bar{K}$ molecule state according to current experiment data. In Ref.~\cite{model9}, the $\Omega(2012)$ resonance was most likely to be the spin-parity $J^{P}=\frac{3}{2}^{-}$ $1P-$wave state, and it had a large potential to be observed in the $\Omega(1672)\gamma$ channel within a nonrelativistic constituent quark potential model. In the hadronic molecular approach, the $\Omega(2012)$ state was considered to contain the mixed $\Xi^{*}\bar{K}$ and $\Omega\eta$ hadronic components~\cite{model10}. A study of the interaction of the $\Xi^{*}\bar{K}$, $\Omega\eta$, and $\Xi\bar{K}$($D-$wave) channels within a coupled channel unitary approach showed that all data including the recent Belle experiment on $\tau_{\Omega^{*}\rightarrow\pi\bar{K}\Xi}/\tau_{\Omega^{*}\rightarrow\bar{K}\Xi}$, were compatible with the molecular picture stemming from meson baryon interaction of these channels~\cite{model11}.

Inspired by the experiment results of the $\Omega(2012)$ by the Belle collaboration and a number of theoretical explanations of this state, we take the chiral quark model (ChQM) and the quark delocalization color screening model (QDCSM) to investigate the pentaquark systems composed of $sssu\bar{u}$ to see whether the $\Omega(2012)$ be described as the molecular picture or not. By using the constituent quark models, we have studied the hidden-charm pentaquark systems~\cite{Huang0,Huang1} and predicated several resonance states, three of which were consistent with the reported $P_{c}(4312)$, $P_{c}(4440)$, and $P_{c}(4457)$ by the LHCb collaboration~\cite{LHCb}. Extending to the study of the hidden-strange pentaquark systems, we also obtained several $P_{c}-$like resonance states~\cite{Huang2}. Therefore, it is feasible to apply these models to the $sssu\bar{u}$ systems.

The structure of this paper is as follows. A brief introduction of two quark models is given in section II.
Section III devotes to the numerical results and discussions. The summary is shown in the last section.

\section{Two quark models}

\subsection{Chiral quark  model}
The ChQM is based on the fact that nearly massless current light quark acquires a dynamical, momentum dependent mass, namely, the constituent quark mass due to its interaction with the gluon medium. So the chiral quark model contains Goldstone-boson exchange potentials, the perturbative one-gluon interaction and a linear-screened confining potential. The model details can be found in Ref. ~\cite{Salamanca}. Here only the Hamiltonian is given:
\begin{widetext}
\begin{eqnarray}
H & = & \sum_{i=1}^5\left(m_i+\frac{p_i^2}{2m_i}\right)-T_{CM} +\sum_{j>i=1}^5
\left(V^{C}_{ij}+V^{G}_{ij}+V^{\chi}_{ij}+V^{\sigma_{a}}_{ij}\right), \\
V^{C}_{ij} & = & -a_{c} \boldsymbol{\lambda}^c_{i}\cdot \boldsymbol{
\lambda}^c_{j} ({r^2_{ij}}+v_{0}), \label{sala-vc} \\
V^{G}_{ij} & = & \frac{1}{4}\alpha_s \boldsymbol{\lambda}^{c}_i \cdot
\boldsymbol{\lambda}^{c}_j
\left[\frac{1}{r_{ij}}-\frac{\pi}{2}\delta(\boldsymbol{r}_{ij})(\frac{1}{m^2_i}+\frac{1}{m^2_j}
+\frac{4\boldsymbol{\sigma}_i\cdot\boldsymbol{\sigma}_j}{3m_im_j})-\frac{3}{4m_im_jr^3_{ij}}
S_{ij}\right] \label{sala-vG} \\
V^{\chi}_{ij} & = & V_{\pi}( \boldsymbol{r}_{ij})\sum_{a=1}^3\lambda
_{i}^{a}\cdot \lambda
_{j}^{a}+V_{K}(\boldsymbol{r}_{ij})\sum_{a=4}^7\lambda
_{i}^{a}\cdot \lambda _{j}^{a}
+V_{\eta}(\boldsymbol{r}_{ij})\left[\left(\lambda _{i}^{8}\cdot
\lambda _{j}^{8}\right)\cos\theta_P-(\lambda _{i}^{0}\cdot
\lambda_{j}^{0}) \sin\theta_P\right] \label{sala-Vchi1} \\
V_{\chi}(\boldsymbol{r}_{ij}) & = & {\frac{g_{ch}^{2}}{{4\pi
}}}{\frac{m_{\chi}^{2}}{{\
12m_{i}m_{j}}}}{\frac{\Lambda _{\chi}^{2}}{{\Lambda _{\chi}^{2}-m_{\chi}^{2}}}}%
m_{\chi} \left\{(\boldsymbol{\sigma}_{i}\cdot
\boldsymbol{\sigma}_{j})
\left[ Y(m_{\chi}\,r_{ij})-{\frac{\Lambda_{\chi}^{3}}{m_{\chi}^{3}}}%
Y(\Lambda _{\chi}\,r_{ij})\right] \right.\nonumber \\
&& \left. +\left[H(m_{\chi}
r_{ij})-\frac{\Lambda_{\chi}^3}{m_{\chi}^3}
H(\Lambda_{\chi} r_{ij})\right] S_{ij} \right\}, ~~~~~~\chi=\pi, K, \eta. \\
V^{\sigma_{a}}_{ij} & = & V_{a_{0}}(
\boldsymbol{r}_{ij})\sum_{a=1}^3\lambda _{i}^{a}\cdot \lambda_{j}^{a}+V_{\kappa}(\boldsymbol{r}_{ij})
\sum_{a=4}^7\lambda_{i}^{a}\cdot \lambda _{j}^{a}+V_{f_{0}}(\boldsymbol{r}_{ij})\lambda _{i}^{8}\cdot \lambda_{j}^{8}+V_{\sigma}(\boldsymbol{r}_{ij})
\lambda _{i}^{0}\cdot \lambda _{j}^{0} \label{sala-su3} \\
V_{k}(\boldsymbol{r}_{ij}) & = & -{\frac{g_{ch}^{2}}{{4\pi }}}
{\frac{\Lambda _{k}^{2}}{{\Lambda _{k}^{2}-m_{k}^{2}}}}%
m_{k}\left[ Y(m_{k}\,r_{ij})-{\frac{\Lambda _{k}}{m_{k}}}%
Y(\Lambda _{k}\,r_{ij})\right] , ~~~~~~\chi=a_{0}, \kappa, f_{0}, \sigma. \\
S_{ij}&=&\left\{ 3\frac{(\boldsymbol{\sigma}_i
\cdot\boldsymbol{r}_{ij}) (\boldsymbol{\sigma}_j\cdot
\boldsymbol{r}_{ij})}{r_{ij}^2}-\boldsymbol{\sigma}_i \cdot
\boldsymbol{\sigma}_j\right\},\\
H(x)&=&(1+3/x+3/x^{2})Y(x),~~~~~~
 Y(x) =e^{-x}/x. \label{sala-vchi2}
\end{eqnarray}
\end{widetext}
Where $\alpha_s$ is the quark-gluon coupling constant.
The coupling constant $g_{ch}$ for chiral field is
determined from the $NN\pi$ coupling constant through
\begin{equation}
\frac{g_{ch}^{2}}{4\pi }=\left( \frac{3}{5}\right) ^{2}{\frac{g_{\pi NN}^{2}%
}{{4\pi }}}{\frac{m_{u,d}^{2}}{m_{N}^{2}}}\label{gch}.
\end{equation}
The other symbols in the above expressions have their usual meanings.

\subsection{Quark delocalization color screening model}
The Hamiltonian of QDCSM is almost the same as that of ChQM but with two modifications~\cite{QDCSM0,QDCSM1}: Firstly, there
is no $\sigma$-meson exchange in QDCSM, and secondly, the screened color confinement is used between quark pairs resident
in different clusters. That is
\begin{equation}
V_{ij}^{C}=\left \{ \begin{array}{ll}
-a_{c}\boldsymbol{\mathbf{\lambda}}^c_{i}\cdot
\boldsymbol{\mathbf{ \lambda}}^c_{j}~(r_{ij}^2+ v_0) &
  \mbox{if \textit{i},\textit{j} in the same} \\
  &  \mbox{baryon orbit} \\
-a_{c}\boldsymbol{\mathbf{\lambda}}^c_{i}\cdot
\boldsymbol{\mathbf{
\lambda}}^c_{j}~(\frac{1-e^{-\mu_{ij}\mathbf{r}_{ij}^2}}{\mu_{ij}}+
v_0) & \mbox{otherwise} \end{array} \right.\label{QDCSM-vc}
\end{equation}
where the color screening constant $\mu_{ij}$ is determined by fitting the deuteron properties, $NN$ scattering
phase shifts and $N\Lambda$, $N\Sigma$ scattering cross sections, with $\mu_{uu}=0.45$, $\mu_{us}=0.19$ and
$\mu_{ss}=0.08$, which satisfy the relation, $\mu_{us}^{2}=\mu_{uu}\mu_{ss}$.

The single particle orbital wave functions in the ordinary quark cluster model are the left and right centered single
Gaussian functions:
\begin{eqnarray}
\phi_\alpha(\boldsymbol {S_{i}})=\left(\frac{1}{\pi
b^2}\right)^{\frac{3}{4}}e^ {-\frac{(\boldsymbol {r}-\frac{2}{5}\boldsymbol
{S_i})^2}{2b^2}},
 \nonumber\\
\phi_\beta(-\boldsymbol {S_{i}})=\left(\frac{1}{\pi
b^2}\right)^{\frac{3}{4}}e^ {-\frac{(\boldsymbol {r}+\frac{3}{5}\boldsymbol
{S_i})^2}{2b^2}} .
 \
\end{eqnarray}
The quark delocalization in QDCSM is realized by writing the single particle orbital wave function as a
linear combination of the left and right Gaussians:
\begin{eqnarray}
{\psi}_{\alpha}(\boldsymbol {S_{i}},\epsilon) &=&
\left({\phi}_{\alpha}(\boldsymbol{S_{i}})
+\epsilon{\phi}_{\alpha}(-\boldsymbol{S_{i}})\right)/N(\epsilon),
\nonumber \\
{\psi}_{\beta}(-\boldsymbol {S_{i}},\epsilon) &=&
\left({\phi}_{\beta}(-\boldsymbol{S_{i}})
+\epsilon{\phi}_{\beta}(\boldsymbol{S_{i}})\right)/N(\epsilon),
\nonumber \\
N(\epsilon)&=&\sqrt{1+\epsilon^2+2\epsilon e^{{-S}_i^2/4b^2}}.
\end{eqnarray}
where $\epsilon(\boldsymbol{S}_i)$ is the delocalization parameter determined by the dynamics of the quark system rather than
adjusted parameters. In this way, the system can choose its most favorable configuration through its own dynamics in a larger
Hilbert space.

The parameters of QDCSM are from our previous work~\cite{Huang2} and the parameters of ChQM can be obtained
by adjusting the baryons and mesons masses. We list all parameters in
Table~\ref{parameters}. The calculated baryon masses and mesons masses in comparison with experimental values are shown in Table~\ref{mass}.
\begin{table}[ht]
\caption{\label{parameters} The parameters of two models:
$m_{\pi}=0.7$fm$^{-1}$, $m_{K}=2.51$fm$^{-1}$,
$m_{\eta}=2.77$fm$^{-1}$, $m_{\sigma}=3.42$fm$^{-1}$,
$m_{a_{0}}=m_{\kappa}=m_{f_{0}}=4.97$fm$^{-1}$,
$\Lambda_{\pi}=4.2$fm$^{-1}$, $\Lambda_{K}=5.2$fm$^{-1}$,
$\Lambda_{\eta}=5.2$fm$^{-1}$, $\Lambda_{\sigma}=4.2$fm$^{-1}$,
$\Lambda_{a_{0}}=\Lambda_{\kappa}=\Lambda_{f_{0}}=5.2$fm$^{-1}$,
$g_{ch}^2/(4\pi)$=0.54, $\theta_p$=$-15^{0}$. }
\begin{tabular}{lcccc}
\hline
              &                         &~~QDCSM~~      &~~ ChQM~~  \\   \hline
              &$b$ (fm)                  &~~0.518       &~~0.518    \\
              & $m_u$ (MeV)              &~~313         &~~313      \\
              & $m_d$ (MeV)              &~~313         &~~313      \\
              & $m_s$ (MeV)              &~~573         &~~573      \\
              &$a_c$ (MeV)               &~~58.03       &~~48.59    \\
              &$v^{qq}_{0}$ (MeV)        &-1.2883       &-0.9610   \\
              &$v^{q\bar{q}}_{0}$ (MeV)  &-0.2012       &-0.0602    \\
              &$\alpha_{uu}$             &~~0.5652      &~~0.8437     \\
              &$\alpha_{us}$             &~~0.5239      &~~0.7583     \\
              &$\alpha_{ss}$             &~~0.4506      &~~0.6197     \\
              &$\alpha_{u\bar{u}}$       &~~1.7930      &~~1.7951     \\
              &$\alpha_{u\bar{s}}$       &~~1.7829      &~~1.7839     \\
              &$\alpha_{s\bar{s}}$       &~~1.5114      &~~1.4300     \\
                                                                       \hline

\hline
\end{tabular}
\end{table}

\begin{table}[ht]
\caption{\label{mass}The masses (in MeV) of the ground baryons and mesons.}
\begin{tabular}{lcccccccc}
\hline
               & ~~$N$~~            & ~~$\Delta$~~ & ~~$\Lambda$~~ & ~~$\Sigma$~~
               & ~~$\Sigma^*$~~     & ~~$\Xi$~~    & ~~$\Xi^*$~~   & ~~$\Omega$~~  \\ \hline
QDCSM          & 939          & 1232       & 1124    & 1238
               & 1360         & 1374       & 1496    & 1642       \\
 ChQM   & 939          & 1232       & 1122    & 1266
               & 1343         & 1396       & 1473    & 1622     \\
 Expt.         & 939          &1232        &1116     &1193
               &1385          &1318        &1533     &1672\\  \hline
               &~~$\phi$~~    &~~$\bar{K}^{*}$~~  &~~$\bar{K}$~~  &~~$\pi$~~ &~~$\rho$~~  &~~$\omega$~~  &~~$\eta$~~  \\
 QDCSM         &1020  &892  &495  &139  &890  &842 & 283  \\
 ChQM  &1020  &892  &495  &139  &891  &788 & 227   \\
  Expt.        &1020  &892  &495  &139  &770  &782 & 547  \\ \hline
\end{tabular}
\end{table}

\section{The results and discussions}
In this work, we systematically investigate the pentaquark systems composed of $sssu\bar{u}$ with the quantum numbers $I=0,~1$ and $J^{P}=\frac{1}{2}^{-}, \frac{3}{2}^{-}, \frac{5}{2}^{-}$. Two kinds of quark models are used here, which are QDCSM and ChQM. Our purpose is to see whether the $\Omega(2012)$ can be explained as the molecular pentaquarks. Moreover, we also attempt to explore if there is any other pentauark state. For the first step, the effective potential of all channels is studied because the attractive potential is necessary for forming a bound state or a resonance. Then, in order to check whether there is any bound state
a dynamic calculation including both the single channel and channel-coupling is carried out. Finally, the scattering process of the corresponding open channels is observed to search for any resonance states.

\subsection{Effective potentials}
The effective potential between two clusters is defined as, $V(S)=E(S)-E(\infty)$, where $E(S)$ is the diagonal matrix element of the Hamiltonian of the system in the generating coordinate with the distance $S$ between two clusters. All the effective potentials in two quark models are shown in Fig. 1 and Fig. 2, respectively.

\begin{figure*}
\epsfxsize=3.5in \epsfbox{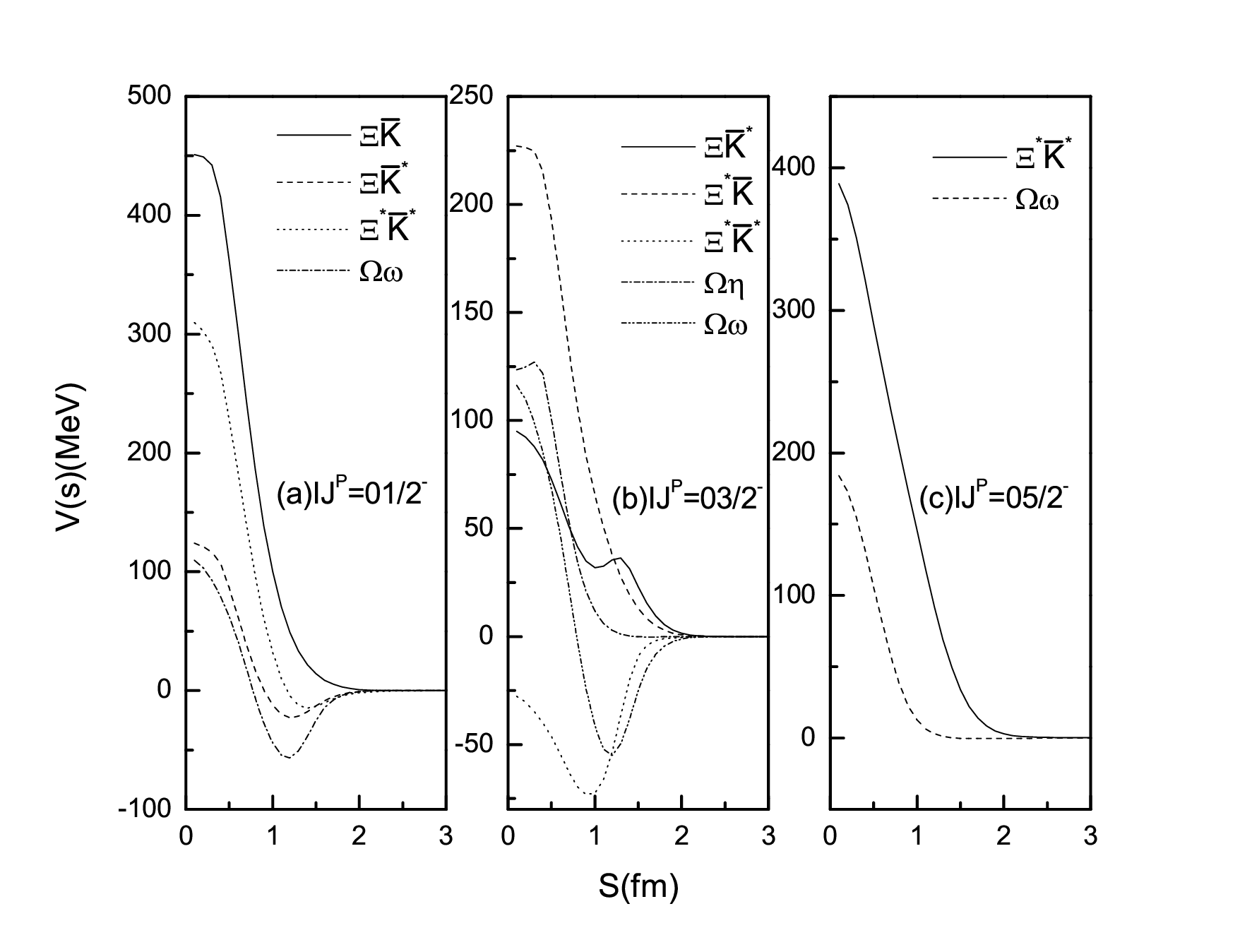} \vspace{-0.2in}
\epsfxsize=3.5in \epsfbox{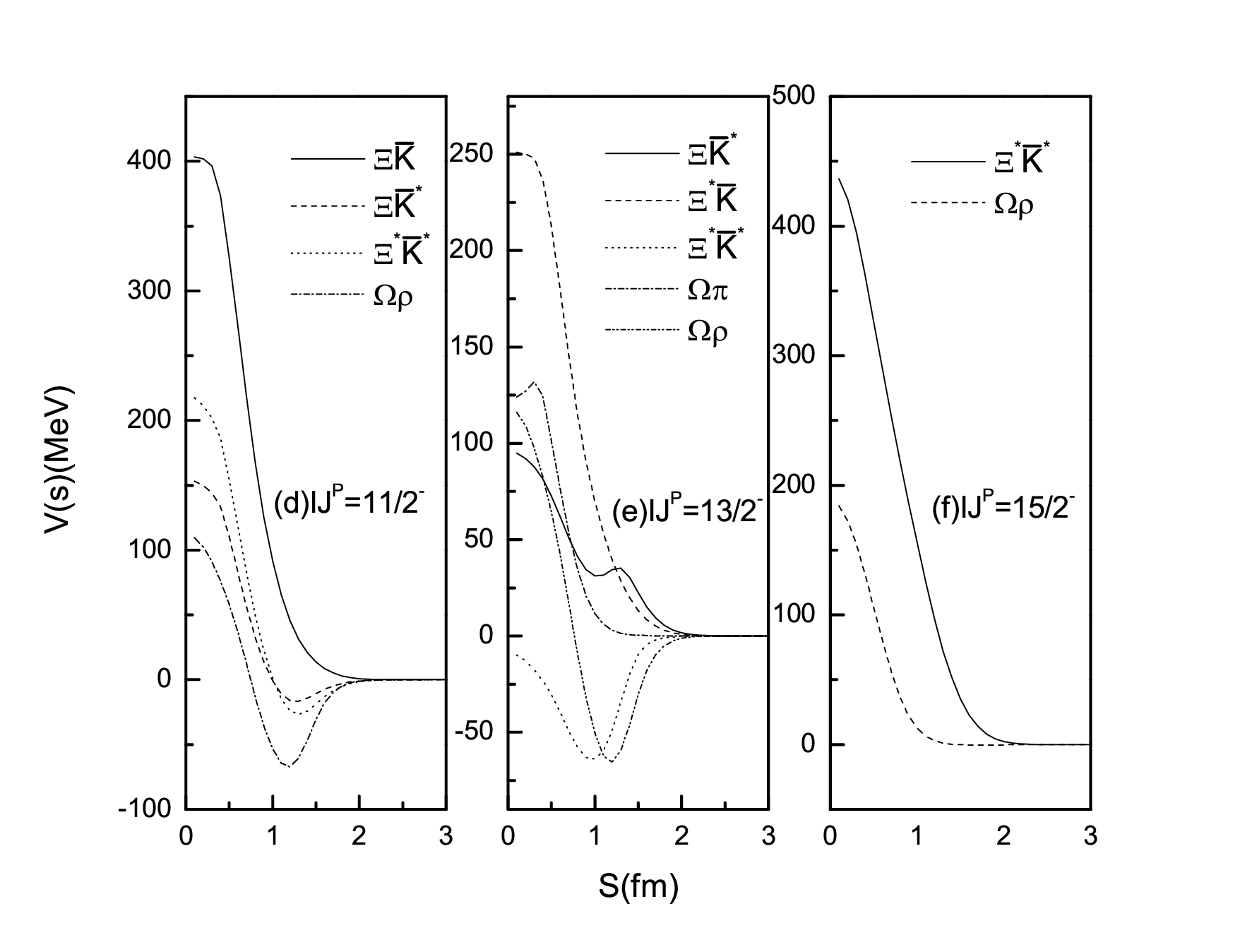} \vspace{-0.2in}
\caption{The effective potentials of the $sssu\bar{u}$ system in QDCSM}
\end{figure*}

\begin{figure*}
\epsfxsize=3.5in \epsfbox{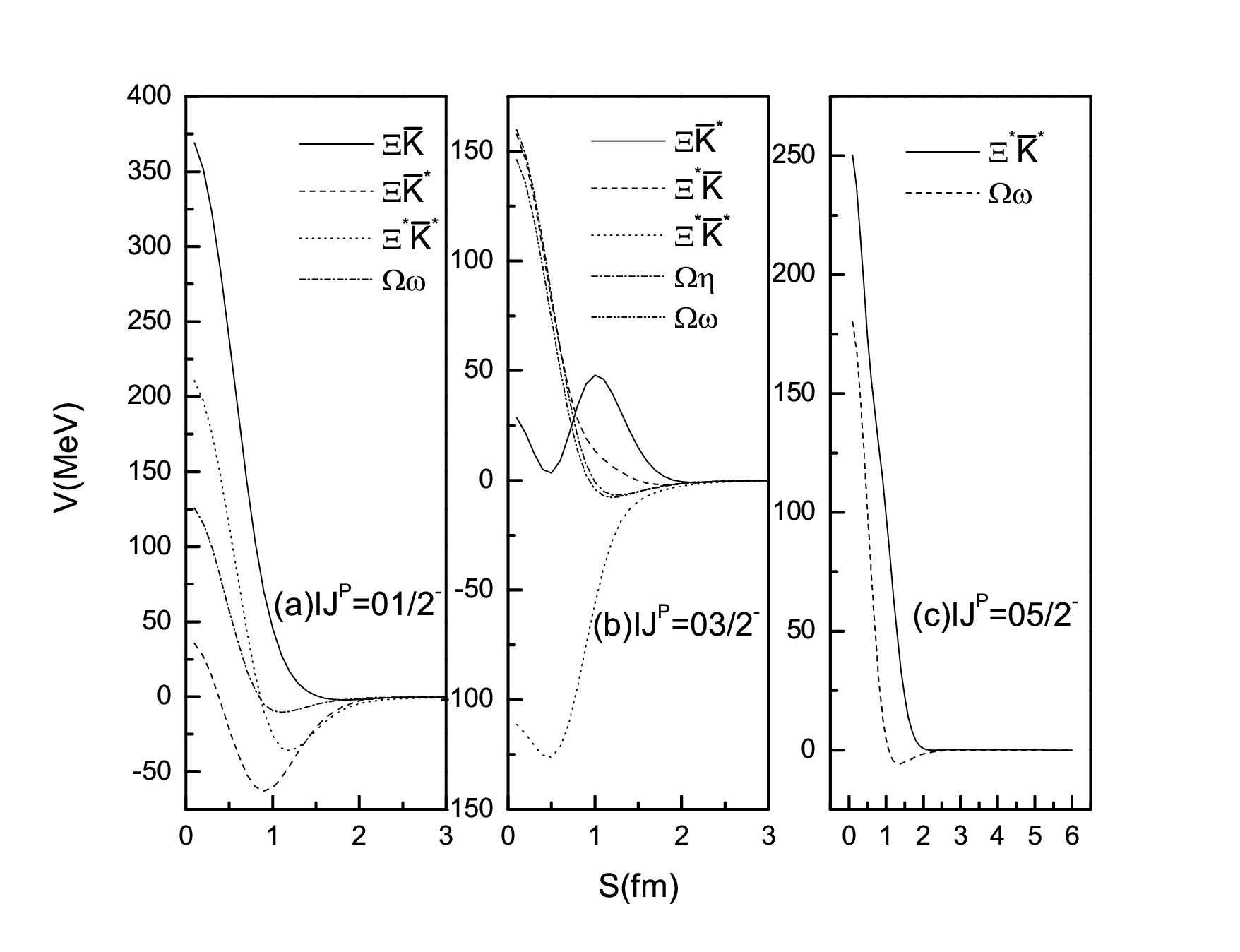} \vspace{-0.2in}
\epsfxsize=3.5in \epsfbox{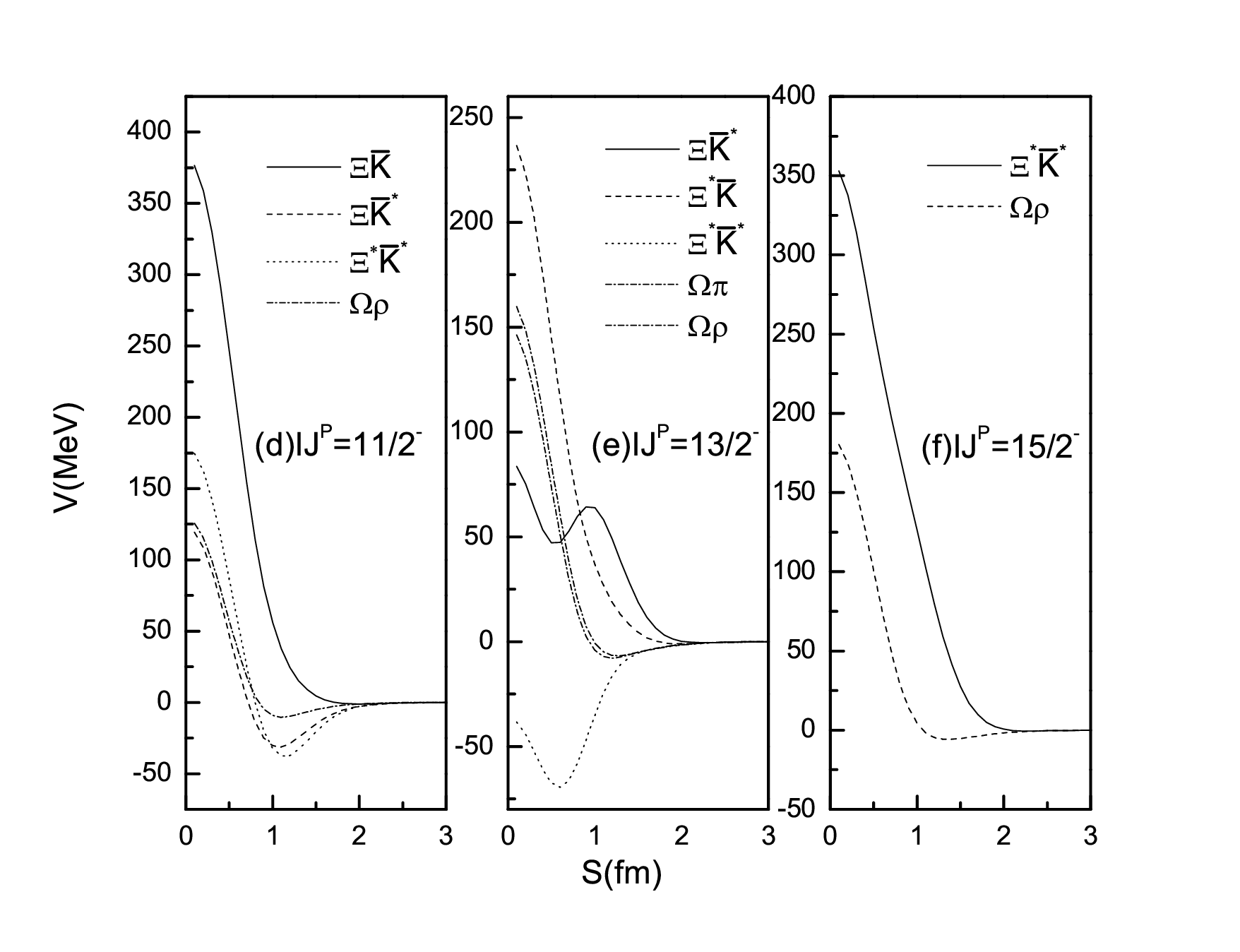} \vspace{-0.2in}
\caption{The effective potentials of the $sssu\bar{u}$ system in ChQM}
\end{figure*}

Fig.1 shows the results in the QDCSM. For the $IJ^{P}=0\frac{1}{2}^{-}$ system, we can see that the potentials are attractive for the $\Omega \omega$, $\Xi^{*}\bar{K}^{*}$, and $\Xi \bar{K}^{*}$ channels, while the one for the $\Xi \bar{K}$ channel is repulsive. It is obvious that the attraction of the $\Omega \omega$ is the largest one, followed by the attractions of $\Xi \bar{K}^{*}$ and $\Xi^{*}\bar{K}^{*}$ channels. So it is more possible for the $\Omega \omega$ channel to form a bound state. For the $IJ^{P}=0\frac{3}{2}^{-}$ system, there are five physical channels, and clearly there is only two channels $\Xi^{*}\bar{K}^{*}$ and $\Omega \omega$, which are attractive.
For the $IJ^{P}=0\frac{5}{2}^{-}$, the system have two physical channels: $\Omega \omega$ and $\Xi^{*}\bar{K}^{*}$, but the interactions of both two channels are repulsive. So it is difficult for these two channels to form any bound state. For the $IJ^{P}=1\frac{1}{2}^{-}$, the potential of the $\Omega\rho$, $\Xi^{*}\bar{K}^{*}$, and $\Xi \bar{K}^{*}$ channels show an attractive property, while the potential of the $\Xi \bar{K}$ channel is repulsive.
For the $IJ^{P}=1\frac{3}{2}^{-}$ and $IJ^{P}=1\frac{5}{2}^{-}$, similar results to that of the $IJ^{P}=0\frac{3}{2}^{-}$ and $IJ^{P}=0\frac{5}{2}^{-}$ are obtained. The potential is attractive for the channel $\Xi^{*}\bar{K}^{*}$ and $\Omega\rho$ with $IJ^{P}=1\frac{3}{2}^{-}$, while the interaction of the rest of channels are all repulsive.

Fig. 2 lists the potentials of all channels in the ChQM. For the $IJ^{P}=0\frac{1}{2}^{-}$ and $IJ^{P}=1\frac{1}{2}^{-}$systems, the behavior of the effective potentials is similar to that of QDCSM. Most channels are attractive, except for the $\Xi \bar{K}$ channel. For the $J^{P}=\frac{3}{2}^{-}$ systems, the behavior of the effective potentials of both $I=0$ and $I=1$ is similar to each other. There are four channels with weak attractions, which are $\Omega \omega$ and $\Omega \eta$ with $IJ^{P}=0\frac{3}{2}^{-}$, and $\Omega \pi$ and $\Omega \rho$ with $IJ^{P}=1\frac{3}{2}^{-}$. Besides, there are two channels with deep attractions, which are the $\Xi^{*}\bar{K}^{*}$ with $IJ^{P}=0\frac{3}{2}^{-}$ and $IJ^{P}=1\frac{3}{2}^{-}$. For the $J^{P}=\frac{5}{2}^{-}$ systems, the potentials of the channels $\Omega\omega$ with $I=0$ and $\Omega \rho$ with $I=1$ are weakly attractive, while the other two channels are repulsive.

According to the results of these two models, we find that the potential of both the $IJ^{P}=0\frac{3}{2}^{-}$ and $IJ^{P}=1\frac{3}{2}^{-}$ $\Xi^{*}\bar{K}^{*}$ channels are attractive in both two quark models, and the attraction is larger than the one of other channels. So, among all the channels of the $sssu\bar{u}$ system, the $\Xi^{*}\bar{K}^{*}$ channel with $IJ^{P}=0\frac{3}{2}^{-}$ and $IJ^{P}=1\frac{3}{2}^{-}$ $\Xi^{*}\bar{K}^{*}$ is the most possible channel to form a bound state.

\subsection{Bound state calculation}
To check whether the possible bound state can be realized, a dynamic calculation is need. Here, the resonating group method(RGM)~\cite{RGM} is employed. Expanding the relative motion wave function between two clusters in the RGM equation by Gaussians, the integro-differential equation of RGM can be reduced to an algebraic equation, the generalized eigenequation. The energy of the system can be obtained by solving the eigenequation. In the calculation, the baryon-meson separation ($|\mathbf{s}_n|$) is taken to be less than 6fm (to keep the matrix dimension manageably small). The masses and binding energies of every single channel and those with channel coupling are listed in Tables~\ref{bound1} and \ref{bound3}, in which $E_{c.c}$ is the results of the channel coupling, and "$\cdots$" means that there is no this channel.

The results of all the systems in the QDCSM are listed in Table~\ref{bound1}. For the $IJ^{P}=0\frac{1}{2}^{-}$ system, the single-channel calculation shows that $\Xi \bar{K}$ is unbound, which agrees with the repulsive nature of the interaction of this channel. For the $\Xi \bar{K}^{*}$ and $\Xi^{*} \bar{K}^{*}$ channels, the attractions are too weak to tie the two particles together, so neither of them is a bound state. However, due to the stronger attraction, the obtained lowest energy of $\Omega \omega$ is below its threshold, with the binding energy of $-4.2$ MeV. We also consider the effect of the multi-channel coupling. The lowest energy by the channel-coupling is $1878.6$ MeV, which is above the threshold of the lowest channel $\Xi \bar{K}$. So there is no bound state for the $IJ^{P}=0\frac{1}{2}^{-}$ system. Nevertheless, we should check if the $\Omega \omega$ is a resonance state in the scattering calculation, which will be presented in the next subsection.

For the $IJ^{P}=0\frac{3}{2}^{-}$, there are two bound states ($\Xi^{*}\bar{K}^{*}$ and $\Omega \omega$ ) by the single calculation. After the multi-channel coupling calculation, the lowest energy of this system is $1864.5$ MeV, $-59.5$ MeV lower than the threshold of the lowest channel $\Omega\eta$, which implies that the $IJ^{P}=0\frac{3}{2}^{-}$ system is bound by the channel coupling. Besides, we also pay attention to the states $\Xi^{*}\bar{K}^{*}$ and $\Omega \omega$, which is possible to be resonance states by the channel coupling. To confirm whether the states $\Xi^{*}\bar{K}^{*}$ and $\Omega \omega$ can survive as resonance states after the full channels coupling, the study of the scatting processes of the open channels is carried out, and the results are shown in the next section. For the $IJ^{P}=0\frac{5}{2}^{-}$, the interaction of two channels is repulsive, so there is no any bound state. The channel coupling calculation cannot help much.

For the $I=1$ system, the behavior is similar to that of the $I=0$ system. There is only one bound state in $IJ^{P}=1\frac{1}{2}^{-}$ system, which is $\Omega \rho$, with binding energy of $-14.1$ MeV. After the channel coupling calculation, the lowest energy of this system is still above the threshold of the $\Xi\bar{K}$ channel, which indicates that this system is unbound by the channel coupling. However, the $\Omega \rho$ may turn out to be a resonance state by coupling to open channels, which should be investigated in the scattering process of the open channels. For the $IJ^{P}=1\frac{3}{2}^{-}$ system, similar results to the case of $IJ^{P}=0\frac{3}{2}^{-}$ system are obtained. The single-channel calculation shows that both the $\Xi^{*}\bar{K}^{*}$ and $\Omega \rho$ are bound state, but both of them can be coupled to the open channels. The scattering process of the open channels will be shown in the next section. Besides, a stable energy is obtained by the channel coupling calculation, the mass of which is $-38.2$ MeV lower than the threshold of $\Omega \pi$. So $IJ^{P}=1\frac{3}{2}^{-}$ system is possible to a bound state after the channel coupling. For the $IJ^{P}=1\frac{5}{2}^{-}$ system, there is no any bound state no matter the single-channel calculation or the channel coupling calculation.


In the ChQM, for the system with $IJ^{P}=0\frac{1}{2}^{-}$, although there are three channels ( $\Xi\bar{K}^{*}$, $\Xi^{*}\bar{K}^{*}$ and $\Omega \omega$), who have attractive potentials, the $\Xi^{*}\bar{K}^{*}$ and $\Omega \omega$ are unbound. It is reasonable. As shown in Fig. 2, the effective potential of $\Xi^{*}\bar{K}^{*}$ and $\Omega \omega$ is weakly attractive, so neither $\Xi^{*}\bar{K}^{*}$ nor $\Omega \omega$ is bound here. However, the attraction between $\Xi$ and $\bar{K}^{*}$ is strong enough to bind $\Xi$ and $\bar{K}^{*}$, so the $\Xi \bar{K}^{*}$ is a bound state with
the binding energy of $-8.1$ MeV in the single calculation. Considering the multi-channel coupling, the results shows that no any bound state exists because the channel coupling cannot push the lowest energy below the threshold of the $\Xi \bar{K}$. However, we should check if $\Xi \bar{K}^{*}$ is a resonance state by coupling the open channel in the following work.
For the $IJ^{P}=0\frac{3}{2}^{-}$ system, since the attraction of the $\Xi^{*}\bar{K}^{*}$ state is the largest one, a bound state with the binding energy of $-58.1$ MeV is obtained in the single channel calculation. However,  the either $\Omega \omega$ or $\Omega \eta$ is bound due to the weak attraction in these two channels. For the $IJ^{P}=0\frac{5}{2}^{-}$ system, the attraction potential between $\Omega$ and $\omega$ is not enough to make a bound state. For the $I=1$ systems, there is only one channel $\Xi^{*}\bar{K}^{*}$ with $J^{P}=\frac{3}{2}^{-}$,  which can form a bound state in a single-channel calculation, because of the strong attraction between $\Xi^{*}$ and $\bar{K}^{*}$ as shown in Fig. 2. Other channels are all unbound, even with the help of multi-channel coupling calculation. However, we still have to investigate the scattering process of the corresponding open channels to confirm if this $\Xi^{*}\bar{K}^{*}$ is a resonance state or not.

\begin{table*}[ht]
\caption{The masses and binding energies of the $sssu\bar{u}$ systems in QDCSM.}
\begin{tabular}{lcccccccccc}
\hline\hline
Channel &
$IJ^{P}$=$0\frac{1}{2}^{-}$ &$IJ^{P}$=$0\frac{3}{2}^{-}$  &$IJ^{P}$=$0\frac{5}{2}^{-}$  &$IJ^{P}$=$1\frac{1}{2}^{-}$ &$IJ^{P}$=$1\frac{3}{2}^{-}$  &$IJ^{P}$=$1\frac{5}{2}^{-}$  &$E_{th}$\\\hline
$\Xi \bar{K}$          &1878.6           &$\cdots$          &$\cdots$ &1878.5              &$\cdots$           &$\cdots$     &1869   \\
$\Xi \bar{K^{*}}$      &2269.3           &2276.1            &$\cdots$ &2271.2              &2275.5             &$\cdots$     &2266  \\
$\Xi^{*} \bar{K^{*}}$  &2394.8           &2365.6/-23.4MeV   &2398.5   &2391.6              &2372.3/-16.7MeV    &2398.3       &2389 \\
$\Xi^{*} \bar{K}$      &$\cdots$         &2000.7            &$\cdots$ &$\cdots$            &2000.4             &$\cdots$     &1991  \\
$\Omega \omega$        &2479.8/-4.2MeV   &2479.7/-4.3MeV    &2539.3   &$\cdots$            &$\cdots$           &$\cdots$     &2484 \\
$\Omega \rho$          &$\cdots$         &$\cdots$          &$\cdots$ &2517.9/-14.1MeV     &2524.6/-7.4MeV     &2539.3       &2532 \\
$\Omega \eta$          &$\cdots$         &1929.8            &$\cdots$ &$\cdots$            &$\cdots$           &$\cdots$     &1924 \\
$\Omega \pi$           &$\cdots$         &$\cdots$          &$\cdots$ &$\cdots$            &1786.8             &$\cdots$     &1780 \\
$E_{c.c}$              &$1878.6$         &1864.5/-59.5MeV   &$2398.1$ &1878.1              &1741.8/-38.2MeV    &$2398.1$     &      \\
\hline\hline
\end{tabular}
\label{bound1}
\end{table*}

\begin{table*}[ht]
\caption{The masses and binding energies of the $sssu\bar{u}$ systems in ChQM.}
\begin{tabular}{lcccccccccc}
\hline\hline
Channel &
$IJ^{P}$=$0\frac{1}{2}^{-}$ &$IJ^{P}$=$0\frac{3}{2}^{-}$  &$IJ^{P}$=$0\frac{5}{2}^{-}$  &$IJ^{P}$=$1\frac{1}{2}^{-}$ &$IJ^{P}$=$1\frac{3}{2}^{-}$  &$IJ^{P}$=$1\frac{5}{2}^{-}$  &$E_{th}$\\\hline
$\Xi \bar{K}$          &1899.5           &$\cdots$          &$\cdots$ &1899.7   &$\cdots$           &$\cdots$     &1891  \\
$\Xi \bar{K^{*}}$      &2279.9/-8.1MeV   &2296.6            &$\cdots$ &2292.3   &2297.3             &$\cdots$     &2288  \\
$\Xi^{*} \bar{K^{*}}$  &2368.1           &2306.9/-58.1MeV   &2374.3   &2368.1   &2351.4/-13.6MeV    &2371.3       &2365  \\
$\Xi^{*} \bar{K}$      &$\cdots$         &1975.4            &$\cdots$ &$\cdots$ &1976.1             &$\cdots$     &1968  \\
$\Omega \omega$        &2417.2           &2417.5            &2417.8   &$\cdots$ &$\cdots$           &$\cdots$     &2410  \\
$\Omega \rho$          &$\cdots$         &$\cdots$          &$\cdots$ &2519.6   &2519.9             &2520.2       &2513  \\
$\Omega \eta$          &$\cdots$         &1856.4            &$\cdots$ &$\cdots$ &$\cdots$           &$\cdots$     &1849  \\
$\Omega \pi$           &$\cdots$         &$\cdots$          &$\cdots$ &$\cdots$ &1768.6             &$\cdots$     &1761  \\
$E_{c.c}$              &$1899.4$         &1856.1            &2374.1   &1899.7   &1768.3             &2371.2       &      \\
\hline\hline
\end{tabular}
\label{bound3}
\end{table*}

Comparing the results of two quark models, although different models obtain several different bound states, some similar conclusions are reached. We find that the interaction between $\Xi^{*}$ and $\bar{K}$ is repulsive, so it is difficult to form a $\Xi^{*} \bar{K}$ bound state. Therefore, the $\Omega(2012)$ cannot be explained as a $J^{P}=\frac{3}{2}^{-}$ $\Xi^{*}\bar{K}$ molecular state in present work. However, the interaction between $\Xi^{*}$ and $\bar{K}^{*}$ is strong enough to form a bound state in both QDCSM and ChQM with $I=0$ or $1$ and $J^{P}=\frac{3}{2}^{-}$, so it is more possible to be a resonance state in the scattering process of the open channels. To check whether or not the $\Xi^{*} \bar{K}^{*}$, as well as other single channel bound states can survive as resonance states after coupling to the open channels, the study of the scattering process of the open channels is performed in the following section.

\subsection{Resonance states}
In this section, we calculate the scattering phase shifts of all the open channels in both QDCSM and ChQM, the results of which are shown in Fig. 3 and Fig. 4, respectively. The energy and decay width of the resonance states can be obtained from the phase shifts, and they are listed in Tables~\ref{decay1},~\ref{decay2}, and \ref{decay3}, where $M_{r}$ is the resonance mass, $\Gamma_{i}$ is the partial decay width of the resonance state, and $\Gamma_{total}$ is the total decay width.
We should mention that all of the states that we study here are in the $S-$wave. The $S-$wave bound states that
decay to $D-$wave open channels through tensor interactions are neglected here due to their small decay widths, so the
total decay widths of the states given below are the lower limits. The general features of the calculated results are as follows.

\begin{figure*}
\epsfxsize=4.0in \epsfbox{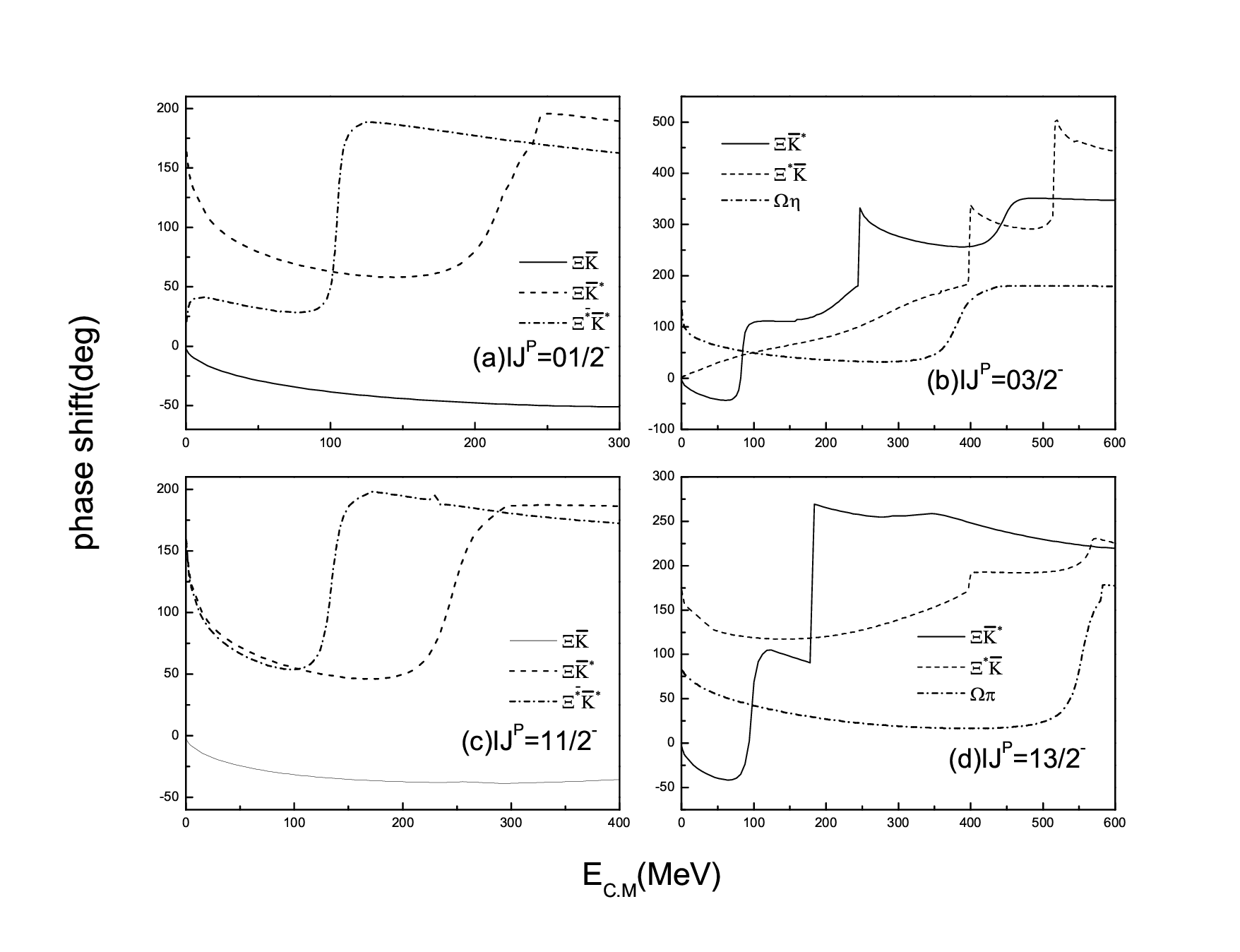} \vspace{-0.2in}
\caption{The phase shifts of the open channels in QDCSM.}
\end{figure*}

\begin{figure}
\epsfxsize=3.5in \epsfbox{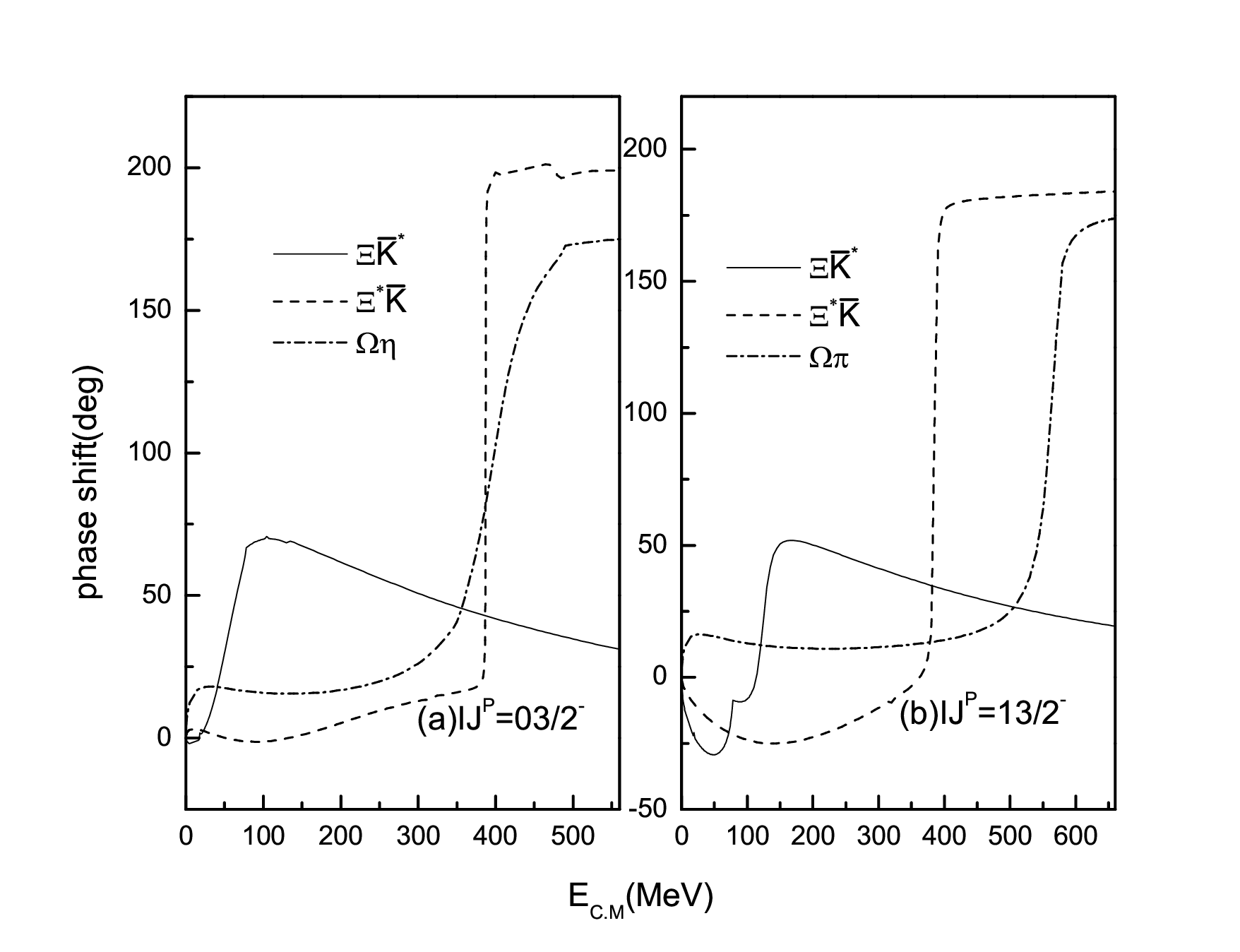} \vspace{-0.3in}
\caption{The phase shifts of the open channels in ChQM.}
\end{figure}

In QDCSM, for the $IJ^{P}=0\frac{1}{2}^{-}$ system, the bound state $\Omega\omega$ can decay to three open channels: the $S-$ wave $\Xi\bar{K}$, $\Xi\bar{K}^{*}$ and $\Xi^{*}\bar{K}^{*}$. The scattering phase shifts of these three open channels are shown in Fig. 3(a), from which we can see that there is no resonance state in the $\Xi\bar{K}$ channel, while the $\Omega\omega$ appears as a resonance state in both $\Xi\bar{K}^{*}$ and $\Xi^{*}\bar{K}^{*}$ channels, with the resonance energy of $2446$ MeV and $2464$ MeV, and the total decay width of $87$ MeV. This indicates that the $\Omega\omega$ with $IJ^{P}=0\frac{1}{2}^{-}$ is possible to be a broad resonance state in the QDCSM.

For the $IJ^{P}=0\frac{3}{2}^{-}$ system, Fig. 3(b) shows the phase shifts of the open channels $\Xi\bar{K}^{*}$, $\Xi^{*}\bar{K}$ and $\Omega\eta$. There are two resonance states in both the $\Xi\bar{K}^{*}$ and $\Xi^{*}\bar{K}$ scattering phase shifts corresponding to $\Xi^{*}\bar{K}^{*}$ and $\Omega\omega$ respectively, while in the $\Omega\eta$ scattering channel there is only one resonance state corresponding to $\Xi^{*}\bar{K}^{*}$. The reason is that the channel coupling pushes the higher state $\Omega\omega$ above the threshold. The resonance masses and decay widths of resonance states are listed in Table~\ref{decay2}, which shows that the $IJ^{P}=0\frac{3}{2}^{-}$ $\Xi^{*}\bar{K}^{*}$ is a resonance state with energy of $2328\sim 2374$ MeV and decay width of $57$ MeV, and $IJ^{P}=0\frac{3}{2}^{-}$ $\Omega\omega$ is a resonance state with energy of $2477\sim 2479$ MeV and decay width of $60$ MeV.

The results of the $IJ^{P}=1\frac{1}{2}^{-}$ system is similar to that of the $IJ^{P}=0\frac{1}{2}^{-}$ system. Fig. 3(c) shows that there is no resonance state in the $\Xi\bar{K}$ channel, while the resonance state $\Omega\rho$ appears clearly in both the $\Xi\bar{K}^{*}$ and $\Xi^{*}\bar{K}^{*}$ channels. The resonance energy and decay width are listed in Table~\ref{decay1}, which are $2500\sim 2511$ MeV and $69$ MeV, respectively.

For the $IJ^{P}=1\frac{3}{2}^{-}$ system, there are two singly bound states $\Omega\rho$ and $\Xi^{*}\bar{K}^{*}$, which can decay to three open channels: the $S-$ wave $\Xi\bar{K}^{*}$, $\Xi^{*}\bar{K}$ and $\Omega\pi$. The phase shifts of these open channels are shown in Fig. 3(d), from which we can see that both $\Omega\rho$ and $\Xi^{*}\bar{K}^{*}$ survive as resonance states in the $\Xi\bar{K}^{*}$ and $\Xi^{*}\bar{K}$ scattering process. However, in the $\Omega\pi$ scattering channel, there is only one resonance state, which is $\Xi^{*}\bar{K}^{*}$ state, since the channel coupling pushes the higher state $\Omega\rho$ above its threshold. From Table~\ref{decay2}, we can see that the resonance energy of $IJ^{P}=1\frac{3}{2}^{-}$ $\Xi^{*}\bar{K}^{*}$ is $2342\sim 2386$ MeV and decay width is $100$ MeV, and the resonance energy of $IJ^{P}=1\frac{3}{2}^{-}$ $\Omega\rho$ is $2522\sim 2528$ MeV and decay width is $99$ MeV.

With regards to the ChQM, for the $IJ^{P}=0\frac{1}{2}^{-}$ system, we study the scattering process of the open channel $\Xi \bar{K}$, and we do not find any resonance state. The reason is that the channel coupling to the $\Xi \bar{K}$ pushes the energy of $\Xi \bar{K}^{*}$ above its threshold. For the $IJ^{P}=0\frac{3}{2}^{-}$ system, the singly bound state $\Xi^{*} \bar{K}^{*}$ can be coupled to three open channels: $\Xi \bar{K}^{*}$, $\Xi^{*} \bar{K}$ and $\Omega\eta$, the scattering phase shifts of which are shown in Fig. 4(a). It is obvious in Fig. 4(a) that the $\Xi^{*} \bar{K}^{*}$ appears as a resonance state in both $\Xi^{*} \bar{K}$ and $\Omega\eta$ scattering process, while in the $\Xi \bar{K}^{*}$ channel, there is no resonance states because the channel coupling pushes the higher state $\Xi^{*} \bar{K}^{*}$ above its threshold.

The results of the $IJ^{P}=1\frac{3}{2}^{-}$ system is similar to that of the $IJ^{P}=0\frac{3}{2}^{-}$ system. The phase shifts of three open channels are shown in Fig. 4(b), from which we can see that the singly bound state $\Xi^{*} \bar{K}^{*}$ appears as a resonance state in both $\Xi^{*} \bar{K}$ and $\Omega\pi$ phase shifts, while in the $\Xi \bar{K}^{*}$ channel, there is no resonance states.

The resonance energy and decay width of the $\Xi^{*} \bar{K}^{*}$ with $I=0$ and $1$, and $J^{P}=\frac{3}{2}^{-}$ are listed in Table~\ref{decay3}, from which we can see that the resonance energy of the $IJ^{P}=0\frac{3}{2}^{-}$ $\Xi^{*} \bar{K}^{*}$ is $2337\sim2352$ MeV and decay width is $65.5$ MeV, and those of the $IJ^{P}=1\frac{3}{2}^{-}$ $\Xi^{*} \bar{K}^{*}$ is $2341\sim2350$ MeV and decay width is $31.5$ MeV.

\begin{table}[ht]
\caption{The resonance mass and decay width (in MeV) of the molecular pentaquarks with $J^{P}=\frac{1}{2}^{-}$ in QDCSM.}
\setlength{\tabcolsep}{0.1mm}{
{\begin{tabular}{cccccc} \hline\hline
\multicolumn{1}{c}{} &\multicolumn{2}{c}{$\Omega\omega(I=0)$} &\multicolumn{1}{c}{} &\multicolumn{2}{c}{$\Omega\rho(I=1)$}  \\
\cline{2-3}\cline{5-6}
~~~{open channels}~~~ & ~~~$M_{r}$~~~ & ~~~$\Gamma_{i}$~~~ & & ~~~$M_{r}$~~~ & ~~~$\Gamma_{i}$~~~ \\ \hline
 {$\Xi\bar{K}^{*}$}      & 2446  & 80     &      &2500   &30    \\
 {$\Xi^{*}\bar{K}^{*}$}  & 2464  & 7      &      &2511   &39    \\
 {$\Xi\bar{K} $}         & $-$   & $-$    &      &$-$    &$-$  \\
 {$\Gamma_{total}$}      &  ~~   & 87     &~~~   &~~~    &69    \\
  \hline\hline
\end{tabular}
\label{decay1}}}
\end{table}

\begin{table}[ht]
\caption{The resonance mass and decay width (in MeV) of the molecular pentaquarks with $J^{P}=\frac{3}{2}^{-}$ in QDCSM.}
\setlength{\tabcolsep}{0.1mm}{
{\begin{tabular}{cccccc} \hline\hline
\multicolumn{1}{c}{} &\multicolumn{2}{c}{$\Omega\omega(I=0)$} &\multicolumn{1}{c}{} &\multicolumn{2}{c}{$\Xi^{*}\bar{K}^{*}(I=0)$} \\
\cline{2-3}\cline{5-6}
~~~{open channels}~~~ & ~~~$M_{r}$~~~ & ~~~$\Gamma_{i}$~~~ & & ~~~$M_{r}$~~~ & ~~~$\Gamma_{i}$~~~ \\ \hline
 {$\Xi\bar{K}^{*}$}  & 2479 & 39      &     &2333   &15       \\
 {$\Xi^{*}\bar{K}$}  & 2477 & 21      &     &2374   &0.001    \\
 {$\Omega \eta$}     & $-$  &$-$      &     &2328   &42       \\
 {$\Gamma_{total}$}  & ~~   &  60     &~~~  &~~~    & 57      \\ \hline
\multicolumn{1}{c}{}&\multicolumn{2}{c}{$\Omega\rho(I=1)$}&\multicolumn{1}{c}{}&\multicolumn{2}{c}{$\Xi^{*}\bar{K}^{*}(I=1)$}     \\
\cline{2-3}\cline{5-6}
 ~~~{open channels}~~~ & ~~~$M_{r}$~~~ & ~~~$\Gamma_{i}$~~~ & & ~~~$M_{r}$~~~ & ~~~$\Gamma_{i}$~~~\\ \hline
 {$\Xi\bar{K}^{*}$}   & 2522  & 27     &     &2342   &20   \\
 {$\Xi^{*}\bar{K}$}   & 2528  & 72     &     &2386   &50    \\
 {$\Omega \pi$}       & $-$   &$-$     &     &2362   &30       \\
 {$\Gamma_{total}$}   &       & 99     &     &       &100     \\
  \hline\hline
\end{tabular}
\label{decay2}}}
\end{table}

\begin{table}[ht]
\caption{The resonance mass and decay width (in MeV) of the molecular pentaquarks with $J^{P}=\frac{3}{2}^{-}$ in CHQM.}
{\begin{tabular}{cccccc} \hline\hline
\multicolumn{1}{c}{} &\multicolumn{2}{c}{$\Xi^{*}\bar{K}^{*}(I=0)$} &\multicolumn{1}{c}{} &\multicolumn{2}{c}{$\Xi^{*}\bar{K}^{*}(I=1)$} \\
\cline{2-3}\cline{5-6}
~~~{open channels}~~~ & ~~~$M_{r}$~~~ & ~~~$\Gamma_{i}$~~~  & & ~~~$M_{r}$~~~ & ~~~$\Gamma_{i}$~~~\\
\hline
 {$\Xi^{*}\bar{K}$}   & 2352  &0.5   &   &2350   &1.5     \\
 {$\Omega \eta$}      & 2337  &65    &   & $-$   &$-$     \\
 {$\Omega \pi$}       & $-$   &$-$   &   &2341   &30      \\
 {$\Gamma_{total}$}   & ~~    &65.5  &   &~~~    &31.5    \\
  \hline\hline
\end{tabular}
\label{decay3}}
\end{table}

\section{Summary}
In this work, we systematically investigate the pentaquark systems composed of $sssu\bar{u}$ in framework of two kinds of quark models, which are QDCSM and ChQM. Our purpose is to see whether the $\Omega(2012)$ can be explained as the molecular pentaquarks. Moreover, we also attempt to explore if there is any other pentauark state. Our results shows that:

(1) Although some work considered the $\Omega(2012)$ as a $J^{P}=\frac{3}{2}^{-}$ $\Xi^{*}\bar{K}$ molecular state~\cite{model4,model5,model6} because its mass is very close to the $\Xi^{*} \bar{K}$ threshold, we find that the interaction between $\Xi^{*}$ and $\bar{K}$ is repulsive, which leads to the difficulty to form a $\Xi^{*} \bar{K}$ bound state. Therefore, the $\Omega(2012)$ cannot be explained as a $J^{P}=\frac{3}{2}^{-}$ $\Xi^{*}K$ molecular state in present work.
Ref.~\cite{yang} investigated the $1S$, $2S$, $1P$ and $2P$ states of light baryons within the chiral quark model, and found that
the baryons could not be used to explain the $\Omega(2012)$ well. Therefore, the $\Omega(2012)$ is more possible to be a mixture state of the three-quark system and five-quark system. An unquenched quark model can be used to study this state, which is our next work.

(2) We also find several bound states and resonance states in present work. The $sssu\bar{u}$ systems with both $IJ^{P}=0\frac{3}{2}^{-}$ and $IJ^{P}=1\frac{3}{2}^{-}$ are bound by the multi-channel coupling in QDCSM. We also obtain four resonance states in QDCSM, which are $\Omega \omega$ with $IJ^{P}=0\frac{1}{2}^{-}$ ($M=2446\sim2464$ MeV, $\Gamma=87$ MeV) and $IJ^{P}=0\frac{3}{2}^{-}$ ($M=2477\sim2479$ MeV, $\Gamma=60$ MeV), and $\Omega\rho$ with $IJ^{P}=1\frac{1}{2}^{-}$ ($M=2500\sim2511$ MeV, $\Gamma=69$ MeV) and $IJ^{P}=1\frac{3}{2}^{-}$ ($M=2522\sim2528$ MeV, $\Gamma=99$ MeV). However, all these states cannot be obtained in ChQM. So more evidence should be explored to confirm the existence of these states. Nevertheless, the $\Xi^{*}\bar{K}^{*}$ with $IJ^{P}=0\frac{3}{2}^{-}$ ($M=2328\sim2374$ MeV, $\Gamma=57\sim65.5$ MeV) and $IJ^{P}=1\frac{3}{2}^{-}$ ($M=2341\sim2386$ MeV, $\Gamma=31.5\sim100$ MeV) are verified to be resonance states in both QDCSM and ChQM. So these two resonance states are worthy of being searched by future experiments.

\acknowledgments{This work is supported partly by the National Natural Science Foundation of China under
Contract No. 11675080, No. 11775118, No. 11535005, and No. 11775050.}

\end{document}